\documentclass{wccm2024}

\usepackage{graphicx}
\usepackage{amsmath}
\usepackage{amsfonts}
\usepackage{amssymb}
\usepackage{times}
\usepackage{cleveref}
\usepackage{braket}
\usepackage{breqn}
\usepackage{natbib}
\usepackage{subcaption}

\title{UNITARY QUANTUM ALGORITHM FOR THE LATTICE-BOLTZMANN METHOD}

\author{DAVID WAWRZYNIAK$^{1}$, JOSEF WINTER$^{2}$ STEFFEN SCHMIDT$^{2}$, THOMAS INDINGER$^{2}$, UWE SCHRAMM$^{3}$, CHRISTIAN JAN{\ss}EN$^{3}$, AND NIKOLAUS A. ADAMS$^{1,2}$}

\heading{David Wawrzyniak, Josef Winter, Steffen Schmidt, Thomas Indinger, Uwe Schramm, Christian Janssen, and Nikolaus A. Adams}

\address{$^{1}$ Technical University of Munich \\
Munich Institute of Integrated Materials, Energy and Process Engineering (MEP) \\
Lichtenbergstr. 4a, 85748 Garching, Germany \\
david.wawrzyniak@tum.de 
\and
$^{2}$ Technical University of Munich \\
Chair of Aerodynamics and Fluid Mechanics \\
Boltzmannstr. 15, 85748 Garching, Germany 
\and
$^{3}$ Altair Engineering Inc. \\
1820 E. Big Beaver Road, Troy, 48083, Michigan, United States
}


\abstract{We present a quantum algorithm for computational fluid dynamics based on the Lattice-Boltzmann method. Our approach involves a novel encoding strategy and a modified collision operator, assuming full relaxation to the local equilibrium within a single time step. 
Our quantum algorithm enables the computation of multiple time steps in the linearized case, specifically for solving the advection-diffusion equation, before necessitating a full state measurement.
Moreover, our formulation can be extended to compute the non-linear equilibrium distribution function for a single time step prior to measurement, utilizing the measurement as an essential algorithmic step. 
However, in the non-linear case, a classical postprocessing step is necessary for computing the moments of the distribution function.
We validate our algorithm by solving the one dimensional advection-diffusion of a Gaussian hill. 
Our results demonstrate that our quantum algorithm captures non-linearity.
}
\begin{document}
\thispagestyle{empty}

\section{INTRODUCTION}

Interest in quantum hardware and quantum algorithms is increasing steadily in recent years. 
Leveraging the exponential scaling capabilities and unique quantum phenomena such as entanglement and interference effects, quantum algorithms show promise in speeding up digital algorithms or facilitating simulations that are infeasible using traditional methods. 
Computational fluid dynamics stands out as a potential candidate, to harness the properties offered by quantum systems.
Martin Kiffner and Dieter Jaksch \citep{kiffner2023tensor} propose a quantum-inspired approach to solve the incompressible Navier-Stokes Equations (NSE) for a lid-driven cavity using tensor networks. 
Their approach demonstrates reductions in both memory usage and runtime.
Pfeffer et al. \citep{pfeffer2022hybrid} present a hybrid quantum-classical reservoir computing model to simulate the nonlinear chaotic dynamics of Lorenz-type models for two-dimensional thermal convection flow.
Lubasch et al. \citep{lubasch2020variational} design variational quantum algorithms to solve both linear and nonlinear partial differential equations with potential use in CFD. 
The algorithm introduced by Gaitan \citep{gaitan2020finding} addresses the numerical solution of the NSE governing a steady-state, inviscid, one-dimensional, compressible fluid flow within a convergent-divergent nozzle utilizing a quantum ordinary differential equation algorithm \citep{kacewicz2006almost}. 
Oz et al. \citep{oz2022solving} subsequently applied a quantum algorithm, building upon the algorithm of Gaitan, to solve the Burgers equation.

Another established approach for solving the NSE is the Lattice-Boltzmann method.
The Lattice-Boltzmann method relies on a lattice formulation of the Boltzmann equation, employing discrete velocity sets to approximate the evaluation of single-particle distribution functions.
Todorova and Steijl \citep{todorova2020quantum} introduced the first quantum algorithm designed to solve the collisionless Boltzmann equation.
Schalkers and Moeller \citep{SCHALKERS2024112816} introduced a quantum algorithm for the collisionless Lattice-Boltzmann equation, employing the quantum Fourier transform for the streaming step and implementing specular reflection boundary conditions.
This first quantum algorithm for the Lattice-Boltzmann method solving the linear advection-diffusion equation was proposed by Budinski \citep{budinski2021quantum}
In a subsequent study, Budinski \citep{ljubomir2022quantum} expanded his approach including the vorticity-stream function formulation for the linear advection-diffusion equation.
In both approaches, the requirement for a complete state measurement and subsequent reinitialization following the computation of one time step is necessary.
Sanavio and Succi \citep{sanavio2024lattice} proposed a quantum algorithm for the non-linear component of the Lattice-Boltzmann method, incorporating the Carleman linearization method to approximate the non-linearity.

We introduce a novel methodology for encoding and executing the lattice-Boltzmann method as a quantum algorithm. 
Initially, we encode the square root of the density field into the probability amplitudes of the state vector. 
The uncomputation of this square root naturally arises during the measurement process, where the likelihood of measuring a state is determined by the square of the probability amplitudes.
Our novel collision operator, applicable to both linear and non-linear scenarios, assumes $\Delta t / \tau = 1$. 
This enables, in the linear case, the utilization of our modified collision operator for computing multiple time steps without necessitating a full state measurement after each individual time step. 
In the non-linear case, we propose a hybrid quantum algorithm wherein only the computation of the moments of the distribution functions is conducted as a classical postprocessing step, while the full non-linearity is addressed within the quantum algorithm.
Despite the requirement for a full state measurement after each time step, we are able to represent the non-linear equilibrium distribution function without resorting to approximation techniques.
For the streaming step, we employ established quantum streaming algorithms.

This paper is structured as follows. 
In \cref{sec.2}, we introduce the Lattice-Boltzmann method and the equations to be solved. 
In \cref{sec.3}, we introduce our novel encoding approach and the collision algorithm for both the linear and non-linear cases of the equilibrium distribution function.
In \cref{sec.4}, we present solutions computed using our quantum Lattice-Boltzmann algorithm, and in \cref{sec.5}, we provide a conclusion based on the findings presented in this work.

\section{METHODOLOGY}
\label{sec.2}
The Lattice-Boltzmann method solves the macroscopic Navier-Stokes equations by discretized single-particle distribution functions $f_i(\mathbf{x},t)$, 
describing the motion of microscopic particles on a mesoscopic scale at point $\mathbf{x}$ and time $t$. 
The evolution equation of the distribution functions is described by the Lattice-Boltzmann equation
\begin{equation}
    f_i(\mathbf{x}+c_i\Delta t)-f_i(\mathbf{x},t)=\Omega(\mathbf{x},t),
    \label{eq.1}
\end{equation}
where $c_i$ is the particle velocity, $\Delta t$ is the time step size and $\Omega(\mathbf{x},t)$ is the particle collision operator.
The “Bhatnagar-Gross-Krook” (BGK) formulation of the collision operator reads
\begin{equation}
    \Omega_{BGK}(\mathbf{x},t)=-\frac{f_i(\mathbf{x},t)-f_i^{eq}(\mathbf{x},t)}{\tau}\Delta t
    \label{eq.2}
\end{equation}
with $f_i^{eq} (x,t)$ being the equilibrium distribution function, and $\tau$ being the relaxation time. 
We use the simplification $\Delta t/\tau=1$ in the following, which is a scheme proposed by Junk and Raghurama \citep{junk1999new}. 
This simplification facilitates the development of numerical methods for the quantum Lattice-Boltzmann algorithm.
Generalization towards arbitrary values of $\Delta t/\tau$ will be considered in a subsequent step.
Substituting \cref{eq.2} into \cref{eq.1} yields the simplified Lattice-Boltzmann equation
\begin{equation}
    f_i (\mathbf{x}+c_i \Delta t,t+\Delta t)=f_i^{eq} (\mathbf{x},t).
    \label{eq.3}
\end{equation}
The non-linear $f_i^{eq}(x,t)$ is defined as
\begin{equation}
    f_i^{eq}(\mathbf{x},t)=w_i\rho\left(1+\frac{c_{i\alpha} u_\alpha}{c_s^2}+\frac{u_\alpha u_\beta (c_{i\alpha} c_{i\beta}-c_s^2 \delta_{\alpha \beta} )}{2c_s^4}\right),
    \label{eq.4}
\end{equation}
where $\rho$ is the mass density, $c_s$ is the speed of sound in lattice units, $w_i$ is the weight factor, $c_{i\alpha}$ is the microscopic velocity, and $\delta_{\alpha \beta}$ is the Kronecker delta.
For linear advection-diffusion problems the equilibrium distribution function can be further simplified to
\begin{equation}
    f_i^{eq} (\mathbf{x},t)=w_i \rho\left(1+\frac{c_{i\alpha} u_\alpha}{c_s^2}\right),
    \label{eq.5}
\end{equation}
by linearization.
Macroscopic quantities are computed from the moments of the distribution functions, where the zeroth moment yields the mass density
\begin{equation}
    \rho(\mathbf{x},t)=\sum_if_i(\mathbf{x},t),
    \label{eq.6}
\end{equation}
and the first moment the macroscopic velocity
\begin{equation}
    \rho\mathbf{u}(\mathbf{x},t)=\sum_ic_if_i(\mathbf{x},t).
    \label{eq.7}
\end{equation}

\section{QUANTUM ALGORITHM}
\label{sec.3}
The Lattice-Boltzmann method can be generally separated into two main steps. 
The first is the collision step, which, in our case, evaluates the equilibrium distribution function given in \cref{eq.6,eq.7}. 
When embedding the density in the probability amplitudes of the state vector, this step has two major challenges: non-linearity and non-unitarity.
We propose a novel quantum algorithm for solving the linear and non-linear equilibrium distribution function by unitary operator while preserving the exponential scaling of lattice cells with on the qubit number. 
Furthermore, we allow for the computation of several time steps using the linearized equilibrium distribution functions without reinitialization of the quantum states.

\subsection{Linear quantum Lattice-Boltzmann method}
In the following, we consider the linear form of the equilibrium distribution function in equation 5 and a one-dimensional D1Q3 velocity set. 
First, we formulate a modified version of the Lattice-Boltzmann \cref{eq.3}
\begin{equation}
    \sqrt{f_i(\mathbf{x}+c_i\Delta t,t+\Delta t)}=\sqrt{f_i^{eq}(\mathbf{x},t)},
\end{equation}
where we considered the BGK collision operator and $\Delta t/ \tau=1$.
The modified equilibrium distribution is
\begin{equation}
    \sqrt{f_i^{eq}(\mathbf{x},t)}=\sqrt{w_i\left(1+\frac{c_{i\alpha}u_\alpha}{c_s^2}\right)},
\end{equation}
according to \cref{eq.5}.
Therefore, we initialize the square root of the initial mass density $\rho(\mathbf{x},t=0)$ once in the state vector using amplitude encoding techniques. 
The resulting state vector is
\begin{equation}
    \ket{\Psi_0}=\frac{1}{\lVert \rho \rVert}\sum_{k \in [0,2^M-1]} \sqrt{\rho_k}\ket{00}_f\ket{k},
\end{equation}
where $\sqrt{\rho_k}$ is the square root of the density value at a lattice cell, and $2^M$ is the total number of lattice cells. 
We denote the quantum register $\ket{q}_f$ holding the distribution functions with subscript $f$.
The unitary collision operator is applied using a RY-gate, a controlled RY-gate and a CNOT-gate.
For simplicity, we assume a uniform advection velocity, although this is not a requirement for our algorithm.
The RY-gate is defined as
\begin{equation}
    RY =     \begin{bmatrix}
        \cos\left(\theta/2\right) & -\sin\left(\theta/2\right) \\
        \sin\left(\theta/2\right) & \cos\left(\theta/2\right)
    \end{bmatrix}.  
\end{equation}
The arguments for the two RY-gate are computed using
\begin{equation}
    \theta_0 = 2\arccos\left(\sqrt{w_0}\right),
    \label{eq.x1}
\end{equation}
\begin{equation}
    \theta_1 = 2\arccos\left(\sqrt{0.5\left(1+\frac{u}{c_s^2}\right)}\right).
    \label{eq.x2}
\end{equation}
Applying the RY-gate on the first qubit in the $\ket{q}_f$ register yields
\begin{equation}
    \ket{\Psi_1}=\frac{1}{\lVert \rho \rVert}\sum_{k \in [0,2^M-1]} \sqrt{w_0\rho_k}\ket{00}_f\ket{k}+\sqrt{w_{12}\rho_k}\ket{01}_f\ket{k},
\end{equation}
where $\sin({\theta_0/2})=w_{12}=w_1+w_2$ and $w_1=w_2$.
Now applying the second RY-gate on the second qubit in the $\ket{q}_f$ register and conditioned on the first qubit in $\ket{q}_f$ and using the CNOT-gate to rearrange the amplitudes evolves the state vector to

\begin{multline}
    \ket{\Psi_2}=\frac{1}{\lVert \rho \rVert} \sum_{k \in [0,2^M-1]} \sqrt{w_0\rho_k}\ket{00}_f\ket{k}+\sqrt{w_1\left(1+\frac{u}{c_s^2}\right)\rho_k}\ket{01}_f\ket{k} \\+\sqrt{w_2\left(1-\frac{u}{c_s^2}\right)\rho_k}\ket{10}_f\ket{k},
\end{multline}
where we use $\sin(\theta_1/2)=\sqrt{0.5\left(1-u/c_s^2\right)}$ and $\frac{1}{2}w_{12}=w_1=w_2$.
The positive streaming operator \citep{sato2021variational} is defined by
\begin{equation}
    P = \sum_{k \in [0,2^M-1]} \ket{(k+1)\text{mod}(2^M)}\bra{k},
\end{equation}
and the negative streaming operator is defined by 
\begin{equation}
    N = \sum_{k \in [0,2^M-1]}\ket{k}\bra{(k+1)\text{mod}(2^M)}.
\end{equation}
Applying now the streaming operators conditioned on the respective qubits in the $\ket{q}_f$ register produces the state vector
\begin{multline}
    \ket{\Psi_3}=\frac{1}{C} \sum_{k \in [0,2^M-1]} \sqrt{w_0\rho_k}\ket{00}_f\ket{k}+\sqrt{w_1\left(1+\frac{u}{c_s^2}\right)\rho_k}\ket{01}_f\ket{(k+1)\text{mod}(2^M)}\bra{k}\ket{k}\\+\sqrt{w_2\left(1-\frac{u}{c_s^2}\right)\rho_k}\ket{10}_f\ket{k}\bra{(k+1)\text{mod}(2^M)}\ket{k},
\end{multline}
The last step of the computation of one time step is to measure the $\ket{q}_f$ register and reset the register to state $\ket{00}_f$. 
This operation collapses the qubits in the register $\ket{q}_f$ to one of the three possible states, where a collapse in the state $\ket{00}_f$ represents the computation of the distribution function $f_0$, $\ket{01}_f$ represents $f_1$, and $\ket{10}_f$ represents $f_2$. 
It is not necessary to store the outcome of this measurement.
After resetting the states $\ket{01}_f$ or $\ket{10}_f$ to the state $\ket{00}_f$ the next iteration of collision and streaming can be applied. 
In a digital implementation, typically after one time step, \cref{eq.6} is evaluated, and the mass density for the next time step is used to compute the equilibrium distribution function for the collision step. 
Considering the linearized LBM model, the computation of the macroscopic variables is not necessary for every time step and is sufficient to perform a measurement only of the last time step. 
This can be shown for the non-modified D1Q3 linear collision example. 
First, we consider the time evolution of the density field $\rho(\mathbf{x},t)$ for one time step and abbreviate for simplicity the entire LBM time step advancement as operator $U_{LBM}$ 
\begin{equation}
    \rho(\mathbf{x},t+1) = U_{LBM}\rho(\mathbf{x},t).
    \label{eq.19}
\end{equation}
Inserting \cref{eq.6} into \cref{eq.19} yields
\begin{equation}
    \rho(\mathbf{x},t+1) = U_{LBM}f_0(\mathbf{x},t)+U_{LBM}f_1(\mathbf{x},t)+U_{LBM}f_2(\mathbf{x},t),
\end{equation}
where the individual distribution functions $f_i$ depend on $\rho(\mathbf{x},t)$, thus showing that evaluating $\rho(\mathbf{x},t)$ at every time step is not necessary. 
This formulation blows up in dimension for a digital simulation, as for every time step, the number of distribution functions increases. 
This is handled in the quantum algorithm by mid-circuit measurements, which drop off distribution functions and only at most three distribution functions are embedded in the state vector. 
The final addition of all possible distribution functions is performed intrinsically by the measurement of the whole quantum system. 
The probability of measuring a state is given by the square of the absolute value of the probability amplitudes. 
Therefore, the computed square root of the density distribution functions in the amplitudes of the state vector is read out as a state measurement probability of the non-squared distribution function. 
The addition is now performed, as all measurements of the quantum register $\ket{q}_f$ are classically stored in the same label $\ket{00}_f$, because of the reset operation. 
This approach also allows for parallel evaluation on multiple quantum computers, where the evolution of each distribution function is computed on different devices. 
Digital postprocessing is then needed to combine the results from all devices. 

\subsection{Non-linear quantum Lattice-Boltzmann method}
The method presented in the previous subsection can be extended to compute the non-linear equilibrium distribution function in \cref{eq.4}. 
This comes with some restrictions. 
It is only possible to compute one time step at a time, digital post-processing with constant renormalization factors is necessary and digital evaluation of the macroscopic density and velocity in \cref{eq.6,eq.7}.
First, the equilibrium distribution function in \cref{eq.4} is rewritten as
\begin{equation}
    f_i^{eq}=w_i \rho \frac{c_{i\alpha}^2-c_s^2}{2c_s^4}\left(u_\alpha +\frac{c_s^2c_{i\alpha}}{c_{i\alpha}^2-c_s^2}\right)^2+w_i\rho \frac{c_{i\alpha}^2-2c_s^2}{2(c_{i\alpha}^2-c_s^2)}.
\end{equation}
For the D1Q3 case, the equilibrium distribution functions simplify to
\begin{equation}
    f_0^{eq} = w_0 \rho\left(1-\frac{3}{2}u^2\right),
    \label{eq.22}
\end{equation}
\begin{equation}
    f_1^{eq} = 3w_1 \rho\left(u+0.5\right)^2+\rho\frac{1}{4}w_1,
    \label{eq.23}
\end{equation}
\begin{equation}
    f_2^{eq} = 3w_2 \rho\left(u-0.5\right)^2+\rho\frac{1}{4}w_2,
    \label{eq.24}
\end{equation}
where $c_s=\frac{1}{\sqrt{3}}$, $c_0=0$, $c_1=1$, and $c_2=-1$ is used.
\Cref{eq.22,eq.23,eq.24} are now manipulated in the same way as in the linear case.
First the square root of every term separately is taken and the first term in \cref{eq.23,eq.24} are additionally multiplied by a factor of $\frac{1}{\sqrt{4}}$, which is needed for algorithmic reasons.
This factor is accounted for in the renormalization process after every time step.
The equilibrium distribution functions result in
\begin{equation}
    \sqrt{f_0^{eq}} = \sqrt{\rho} \sqrt{w_0\left(1-\frac{3}{2}u^2\right)},
    \label{eq.25}
\end{equation}
\begin{equation}
    \sqrt{f_{1,u}^{eq}} +\sqrt{f_{1,c}^{eq}}=  \sqrt{\rho}\sqrt{\frac{3}{4}w_1}\left(u+0.5\right)+\sqrt{\rho}\sqrt{\frac{1}{4}w_1},
    \label{eq.26}
\end{equation}
\begin{equation}
    \sqrt{f_{2,u}^{eq}} +\sqrt{f_{2,c}^{eq}}= \sqrt{\rho}\sqrt{\frac{3}{4}w_2}\left(u-0.5\right)+\sqrt{\rho}\sqrt{\frac{1}{4}w_2},
    \label{eq.27}
\end{equation}
where we have split the first and second equilibrium distributions into a velocity-dependent part with subscript $u$ and a constant part with subscript $c$.
Assuming an initialized $\sqrt{\rho}$ in the state vector, all collision terms can be computed using RY-gates and the following angles
\begin{equation}
    \theta_0 = 2\arccos\left(\sqrt{w_0}\right),
    \label{eq.28}
\end{equation}
\begin{equation}
    \theta_1 = 2\arcsin\left(\sqrt{\frac{3}{2}}u\right),
    \label{eq.29}
\end{equation}\begin{equation}
    \theta_2 = 2\arccos\left(\sqrt{\frac{1}{4}}\right),
    \label{eq.30}
\end{equation}
\begin{equation}
    \theta_3 = 2\arccos\left(u+0.5\right),
    \label{eq.31}
\end{equation}
\begin{equation}
    \theta_4 = 2\arccos\left(u-0.5\right).
    \label{eq.32}
\end{equation}
We want to note that \cref{eq.28,eq.30} are constant, and in \cref{eq.29,eq.31,eq.32} $u$ is linear and rescaled by a constant factor.
In the following, we show that using \cref{eq.28,eq.29,eq.30,eq.31,eq.32} the \cref{eq.25,eq.26,eq.27} can be realized in the quantum algorithm.
The whole collision block is performed using RY-gates, (multi-)controlled RY-gates, (multi-)controlled X-Gates, and multi-controlled Hadamard gates.
We start with the initialized state vector
\begin{equation}
    \ket{\Psi_0} = \sqrt{\rho_k} \ket{000}_f\ket{k},
\end{equation}
where we here dropped the summation over $k$ and the normalization factor for better readability.
First, we use a RY-gate with $\theta_0$ on the first qubit of the $\ket{q}_f$ register
\begin{equation}
    \ket{\Psi_1} = \left(\sqrt{w_0}\ket{000}_f+\sqrt{1-w_0}\ket{001}_f\right) \sqrt{\rho_k}\ket{k}.
\end{equation}
We can rewrite $1-w_0 = w_{12}=w_1+w_2$, because of the condition $w_0+w_1+w_2=1$.
In the following we assume a uniform velocity $u$ for convenience, although this is not a necessary requirement for the algorithm.
A non-uniform $u$ requires a series of controlled RY-gates, and is for the following demonstration omitted.
Now we apply CNOT-gates to switch the state $\ket{001}_f$ with $\ket{010}_f$ and we apply a controlled RY-gate on the first qubit of the $\ket{q}_f$ register resulting in the state
\begin{dmath}
    \ket{\Psi_2} = \left(\sqrt{w_0\left(1-\frac{3}{2}u^2\right)}\ket{000}_f+\sqrt{w_0\frac{3}{2}}u\ket{001}_f+\sqrt{w_{12}}\ket{010}_f\right) \sqrt{\rho_k}\ket{k},
\end{dmath}
where we here used $\cos(\theta_1/2)=\cos(\arcsin(\sqrt{3/2}u)) = \sqrt{1-(\sqrt{3/2}u)^2}$.
The state $\ket{000}_f$ is the non-linear resting equilibrium distribution function encoded, and $\ket{001}_f$ is a state which is discarded in the measurement process.
Now a controlled RY-gate using $\theta_3$ is applied on the second qubit of the $\ket{q}_f$ register such that
\begin{equation}
    \ket{\Psi_3} = \left(\sqrt{w_0\left(1-\frac{3}{2}u^2\right)}\ket{000}_f+\sqrt{w_0\frac{3}{2}}u\ket{001}_f+\sqrt{\frac{1}{4}w_{12}}\ket{010}_f+\sqrt{\frac{3}{4}w_{12}}\ket{011}_f\right) \sqrt{\rho_k}\ket{k}.
\end{equation}
In the next step, the state vector is rearranged using multicontrolled X-gates, where the state $\ket{011}_f$ is swapped to $\ket{100}_f$.
Now two multicontrolled H-gates are applied on the first qubit, conditioned such that the H-gate is applied on the states $\ket{010}_f$ and $\ket{100}_f$ resulting in
\begin{dmath}
    \ket{\Psi_4} = \left(\sqrt{w_0\left(1-\frac{3}{2}u^2\right)}\ket{000}_f+\sqrt{w_0\frac{3}{2}}u\ket{001}_f+\sqrt{\frac{1}{4}w_{1}}\ket{010}_f+\sqrt{\frac{1}{4}w_{2}}\ket{011}_f+\sqrt{\frac{3}{4}w_{1}}\ket{100}_f+\sqrt{\frac{3}{4}w_{2}}\ket{101}_f\right) \sqrt{\rho_k}\ket{k},
\end{dmath}
where we here used $\frac{1}{2}w_{12}=w_1=w_2$.
Again a rearrangement of states using multicontrolled X-gates is performed, such that $\ket{101}_f$ is switched to $\ket{110}_f$.
Finally, two multicontrolled RY-gates are applied with the argument $\theta_3$ and $\theta_4$, respectively, such that the state vector evolves to
\begin{dmath}
    \ket{\Psi_5} = \left(\sqrt{w_0\left(1-\frac{3}{2}u^2\right)}\ket{000}_f+\sqrt{w_0\frac{3}{2}}u\ket{001}_f+\sqrt{\frac{1}{4}w_{1}}\ket{010}_f+\sqrt{\frac{1}{4}w_{2}}\ket{011}_f+\sqrt{\frac{3}{4}w_{1}}(u+0.5)\ket{100}_f+\sqrt{\frac{3}{4}w_{1}}\sqrt{1-(u+0.5)^2}\ket{101}_f+\sqrt{\frac{3}{4}w_{2}}(u-0.5)\ket{110}_f+\sqrt{\frac{3}{4}w_{2}}\sqrt{1-(u-0.5)^2}\ket{111}_f\right) \sqrt{\rho_k}\ket{k}.
    \label{eq.39}
\end{dmath}
The states $\ket{101}_f$ and $\ket{111}_f$ are also discarded in the measurement process.
Following \cref{eq.39} all terms that are present in \cref{eq.22,eq.23,eq.24} are now embedded in the amplitudes of the state vector.
After that, streaming algorithms are applied, which stream the subspaces of $\ket{010}_f$ and $\ket{100}_f$ in positive direction and $\ket{011}_f$ and $\ket{110}_f$ in negative direction.
Now the full state vector is measured, where the measurement probability is represented by the square of the probability amplitudes, yielding the computation of the non-linear distribution functions.
After renormalization and accounting for the additional factor of $4$ for the states $\ket{100}$ and $\ket{110}_f$, the moments according to \cref{eq.6,eq.7} can be computed in a classical postprocessing step.

\section{Results}\label{sec.4}

The analytical solution of the advection-diffusion of a Gaussian hill with uniform velocity $u$ is given by
\begin{equation}
    \rho(\mathbf{x},t)=\frac{\sigma_0^2}{\sigma_0^2+\sigma_D^2}\rho_0\text{ exp}\left(-\frac{(\mathbf{x}-\mathbf{x}_0-\mathbf{u}t)^2}{2(\sigma_0^2+\sigma_D^2)}\right),
\end{equation}
where $\sigma_D=\sqrt{2Dt}$, and $\rho$ is the density field.
The initial density distribution is 
\begin{equation}
    \rho(\mathbf{x},t=0)=\rho_0 \text{ exp}\left(-\frac{(\mathbf{x}-\mathbf{x}_0)^2}{2\sigma_0^2}\right).
    \label{eq.density_0}
\end{equation}
We set the initial peak, which is located in the middle of the domain, to $\rho_0=0.1$, and add an ambient density value of $0.1$, we set the velocity $u=0.3$, $\sigma_0=4$, $\Delta x = 1$, $\Delta t = 1$, and the D1Q3 velocity set is used.
We assume periodic boundary conditions.

\subsection{Linearized collision}

We simulate the advection-diffusion of the Gaussian defined in Equation \eqref{eq.density_0} for 20 time steps. 
We compare the solutions obtained from the linearized quantum Lattice-Boltzmann method, the digital reference Lattice-Boltzmann method, and the analytical solution.
In our quantum algorithm, we leverage the Qiskit SDK \citep{Qiskit}, utilizing its AER backend for simulating quantum computers. 
Specifically, we employ a shot-based and noise-free simulator provided by the AER backend, utilizing 900,000 shots for our simulation.
An exact agreement between the digital LBM and quantum LBM solutions is obtained. 
Deviations arise primarily from sampling errors inherent in the quantum algorithm, which can be reduced by increasing the number of shots. 
The observed discrepancies in both the quantum and digital LBM solutions compared to the analytical solution are attributed to factors such as spatial and temporal discretization, as well as the linearization of the equilibrium distribution function, which introduces an unintended correction of the diffusivity \citep{chopard2009lattice}.
\subsection{Non-linear collision}

We replicate the scenario from the linearized case in our computation. 
For the quantum algorithm, we utilize the deterministic state vector simulator available in the Qiskit SDK \citep{Qiskit}.
We observe no deviation from the digital non-linear LBM. 
Additionally, the deviation from the analytical solution is reduced, as the deviation of diffusivity for the non-linear collision operator is no longer dependent upon the advection velocity, contrary to the linear case \citep{chopard2009lattice}.

\begin{figure}
    \centering
    \begin{subfigure}{0.45\textwidth}
        \includegraphics[width=\textwidth]{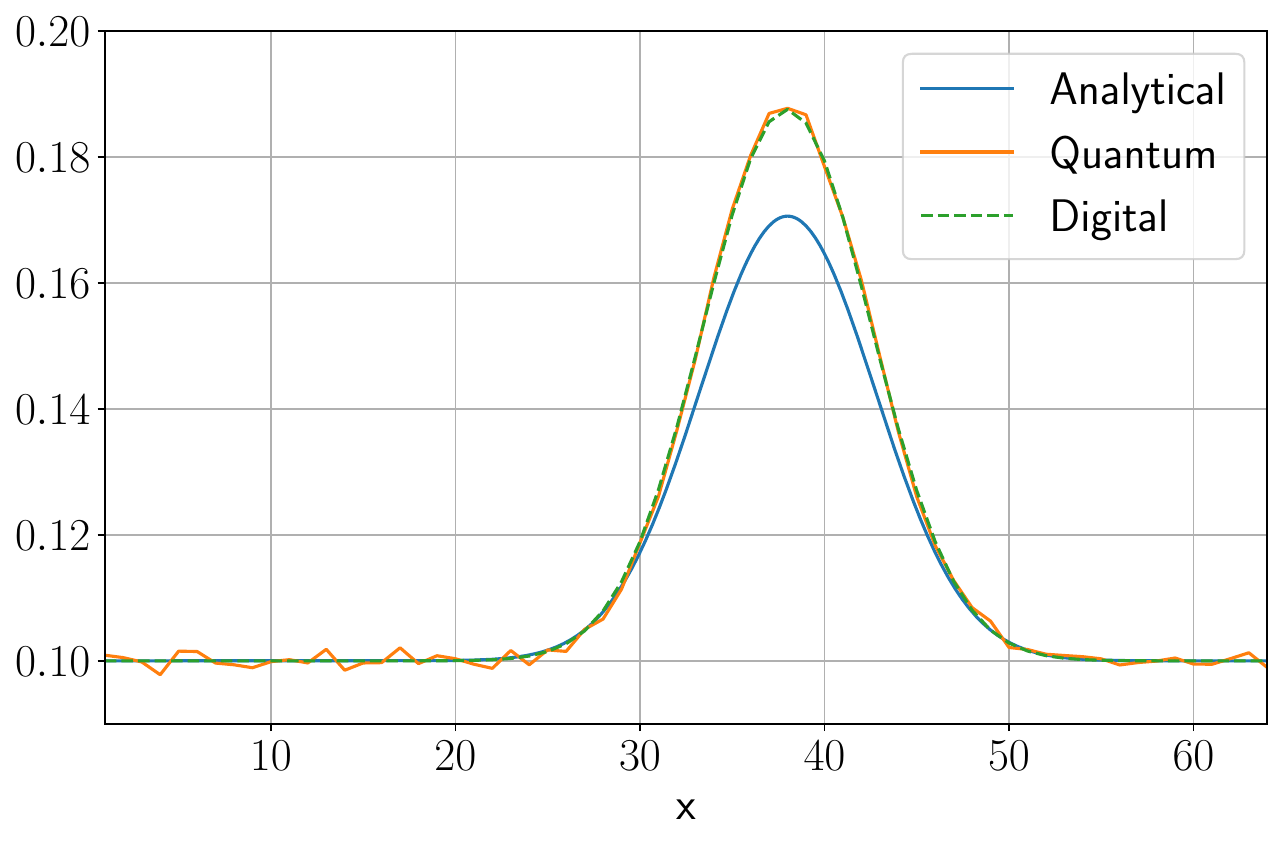}
        \caption{Solving ADE using linearized LBM.}
    \end{subfigure}
    \hfill
    \begin{subfigure}{0.45\textwidth}
        \includegraphics[width=\textwidth]{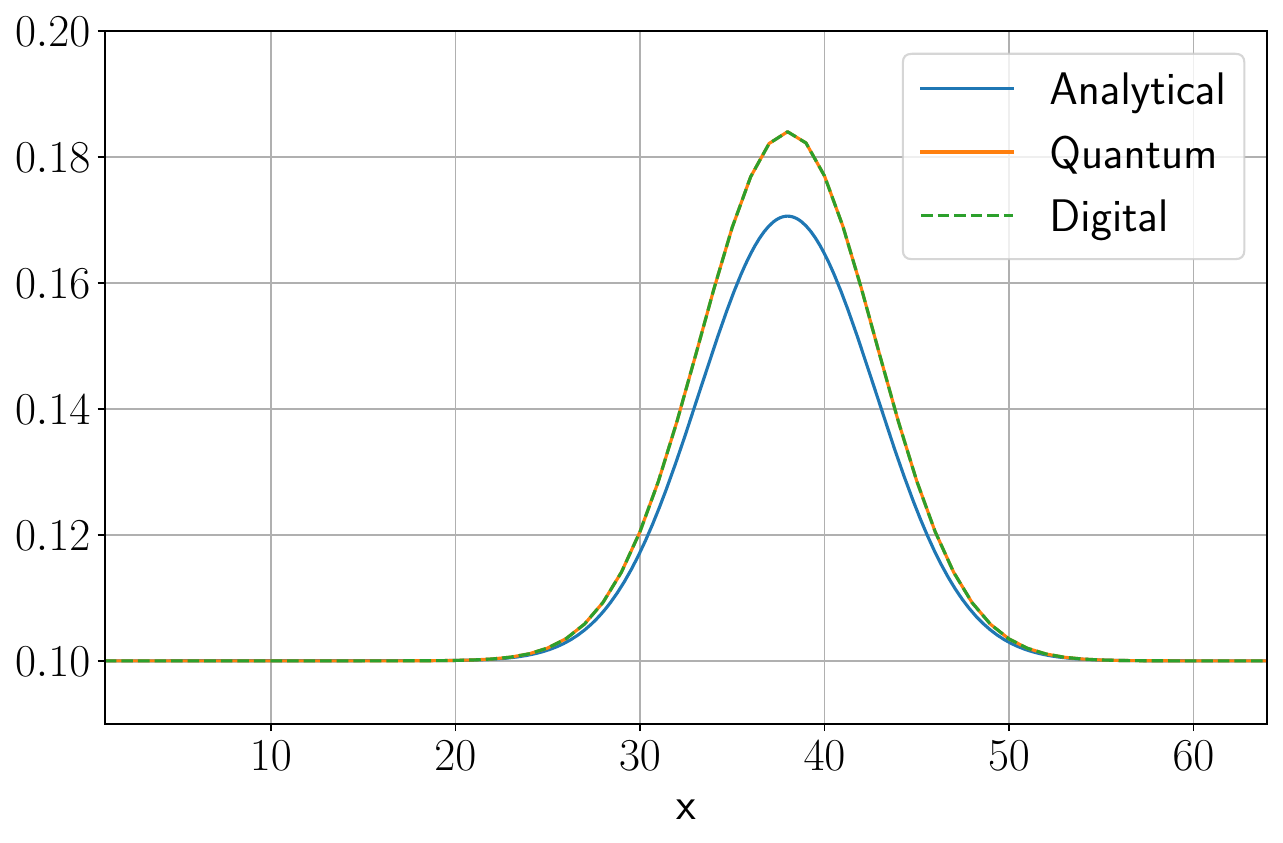}
        \caption{Solving ADE using non-linear LBM.}
    \end{subfigure}
    \caption{Results of the advection-diffusion of a Gaussian hill in 1D. The linearized QLBM is solved using a sampling based simulator and the non-linear QLBM is solved using the state vector simulator.}
    \label{fig:16}
\end{figure}

\section{Conclusion}\label{sec.5}

The contribution of this paper is in introducing a novel encoding approach for the quantum Lattice-Boltzmann method. 
This methodology enables the reliable computation of multiple time steps in the linearized case and facilitates the calculation of the non-linear collision operator for a single time step before conducting a full state measurement.
We have introduced a quantum collision operator tailored for the linearized Lattice-Boltzmann method. 
This operator is adaptable and can be readily customized for more complex velocity sets or additional spatial dimensions as needed.
We have formulated the non-linear collision operator utilizing solely standard RY-gates, (multi-)controlled X-gates, and (multi-)controlled Hadamard-gates. 
We provide all the necessary parameters for the RY-gates, which are computed based on the macroscopic velocity at a lattice site and are adjusted by a constant factor. 
Additionally, all other parameters remain constant for a given velocity set.
Our algorithm lends itself readily to extension for non-uniform velocities. 
This can be achieved by computing the arguments of the RY-gate individually for each velocity on a lattice site and using proper conditioning of the RY-gates.
Similarly, the extension to higher spatial dimensions and more complex velocity sets can be executed analogously, maintaining consistency with the methodology employed for the presented implementation.
We have demonstrated exact agreement between the results obtained from our digital LBM and the quantum LBM.
Future research will concentrate on implementing various boundary conditions and developing methodologies for performing multiple time steps before a measurement is necessary specifically tailored for the non-linear collision operator.

\section*{Acknowledgments}

The authors gratefully acknowledge funding for this research from Altair Engineering Inc.

\bibliographystyle{plainnat}
\bibliography{refs.bib}

\end{document}